\def\lesssim{\mathrel{\hbox{\rlap{\hbox{\lower4pt\hbox{$\sim$}}}\hbox{$<$}}}}
\def\gtrsim{\mathrel{\hbox{\rlap{\hbox{\lower4pt\hbox{$\sim$}}}\hbox{$>$}}}}

\def\ll_lsun{log$({L/\rm L_{\odot}})$~}
\def\masa_msun{$M/ \rm M_{\odot}$~}
\def\m_mstar{$M/M_{*}$~}

\documentclass{aa}
\usepackage{graphicx}
        
\begin{document}

\title{Pulsations  of  massive  ZZ  Ceti stars with carbon/oxygen and 
       oxygen/neon cores}

\author{A. H.    C\'orsico$^{1,2}$\thanks{Member    of    the    
        Carrera    del
        Investigador   Cient\'{\i}fico   y   Tecnol\'ogico,   CONICET,
        Argentina.},    E.     Garc\'{\i}a--Berro$^{3,4}$,   L.     G.
        Althaus$^{1,2,3}$\thanks{Member    of    the    Carrera    del
        Investigador   Cient\'{\i}fico   y   Tecnol\'ogico,   CONICET,
        Argentina.}  \and J. Isern$^{3,5}$}

\offprints{A. H. C\'orsico}

\institute{$^1$Facultad de Ciencias Astron\'omicas y  Geof\'{\i}sicas, 
	   Universidad  Nacional de  La Plata,  Paseo del Bosque, s/n,
	   (1900) La Plata, Argentina.\\
	   $^2$Instituto  de  Astrof\'{\i}sica   La  Plata,  IALP,
	   CONICET\\   
           $^3$Departament   de  F\'\i   sica   Aplicada,  Universitat
	   Polit\`ecnica de Catalunya, Av. del Canal Ol\'\i mpic, s/n,
   	   08860 Castelldefels, Spain\\
	   $^4$Institut d'Estudis Espacials  de Catalunya, Ed.  Nexus,
	   c/Gran Capit\`a 2, 08034 Barcelona, Spain.\\
           $^5$ Institut de Ci\`encies de l'Espai (CSIC)\\
\email{acorsico@fcaglp.unlp.edu.ar, leandro, garcia@fa.upc.es, 
       isern@ieec.fcr.es} }

\date{Received; accepted}

\abstract{We explore the  adiabatic pulsational properties of  massive 
white  dwarf stars  with hydrogen-rich  envelopes and  oxygen/neon and
carbon/oxygen cores.  To  this end, we compute the  cooling of massive
white dwarf models for both  core compositions taking into account the
evolutionary  history  of  the   progenitor  stars  and  the  chemical
evolution caused by  time-dependent element diffusion.  In particular,
for the  oxygen/neon models, we  adopt the chemical  profile resulting
from repeated  carbon-burning shell  flashes expected in  very massive
white dwarf  progenitors.  For carbon/oxygen white  dwarfs we consider
the   chemical   profiles  resulting   from   phase  separation   upon
crystallization. For  both compositions we also take  into account the
effects  of crystallization  on the  oscillation eigenmodes.   We find
that  the  pulsational  properties  of oxygen/neon  white  dwarfs  are
notably  different  from  those  made of  carbon/oxygen,  thus  making
asteroseismological techniques a  promising way to distinguish between
both types of  stars and, hence, to obtain  valuable information about
their progenitors.
 }

\authorrunning{C\'orsico et al.}
\titlerunning{Pulsations of massive ZZ Ceti stars with CO and ONe cores}

\maketitle


\section{Introduction}

White  dwarfs are the  most common  end-product of  stellar evolution.
Most of the presently  observed white dwarfs are post-Asymptotic Giant
Branch (AGB) remnants, the core of which consists of the main ashes of
helium burning,  basically a mixture  of carbon and  oxygen.  However,
there  is growing  theoretical evidence  suggesting that  white dwarfs
with  masses larger  than  $\sim  1.05 \,  M_\odot$  could have  cores
composed primarily of oxygen and neon (Garc\'\i a--Berro \& Iben 1994;
Ritossa et al.  1996;  D'Antona \& Mazzitelli 1996; Garc\'{\i}a--Berro
et al.  1997; Iben et  al. 1997).  In particular, Garc\'{\i}a-Berro et
al.  (1997) found that when the  core mass of the $9 \, M_\odot$ white
dwarf progenitor  exceeds $\sim 1.05 \, M_\odot$,  and before reaching
the  thermally  pulsing  phase  at  the AGB  tip,  carbon  is  ignited
off-center  in semidegenerate  conditions.   As a  result of  repeated
carbon-burning  shell  flashes that  eventually  give  rise to  carbon
exhaustion in the degenerate core, it  is found that at the end of the
carbon burning phase  the star would be left  with an oxygen/neon core
with  trace amounts  of  carbon and  other  heavier chemical  species.
After  considerable  mass-loss  episodes,  the progenitor  remnant  is
expected to  evolve into  the central star  of a planetary  nebula and
ultimately into a  white dwarf with an oxygen/neon  core of mass $\sim
1.05  \,  M_\odot$.  A  possible  observational  counterpart of  these
massive white dwarfs would be the single massive white dwarf LHS~4033,
which has a mass of $\sim  1.32\, M_\odot$ (Dahn et al.  2004).  Other
possible massive  white dwarfs hosting oxygen/neon cores  would be the
magnetic white dwarf PG~1658+441 (Schmidt  et al.  1992; Dupuis et al.
2003) --- with a  mass of $\simeq  1.31\, M_\odot$ --- and  the highly
magnetic white dwarf RE~J0317--853  (Barstow et al.  1995; Ferrario et
al. 1997),  which has a  mass of $\sim  1.35\, M_\odot$. It  should be
noted, however, that there  are alternative evolutionary channels that
could  eventually  lead to  massive  white  dwarfs with  carbon/oxygen
cores.   These   scenarios  involve  the  merging   of  two  otherwise
light-weight ordinary  white dwarfs. It  has been shown  (Segretain et
al. 1997;  Guerrero et al. 2004)  that in this  case nuclear reactions
are not able to modify considerably the composition of the core of the
remnant.

Many white  dwarfs exhibit multiperiodic  luminosity variations caused
by gravity-modes  of low harmonic  degree ($\ell \leq 2$)  and periods
ranging from  roughly 100~s to approximately 1200~s.   Over the years,
the study of the pulsational patterns of variable white dwarfs through
asteroseismological  techniques has  become a  very powerful  tool for
probing  the inner  regions that  would be  otherwise  unaccessible to
direct  observations.  In  particular, for  hydrogen-rich  variable DA
white dwarfs  --- also known as  ZZ Ceti stars ---  such technique has
proved  to  be  very  successful  in  providing  independent  valuable
constraints to  their fundamental  properties.  Examples of  these are
their core composition, the outer layer chemical stratification or the
stellar  mass ---  see, for  instance,  Pfeiffer et  al.  (1996),  and
Bradley  (1998,  2001),  amongst  others.  Very  recently,  increasing
attention has been paid to study the asteroseismological properties of
massive ZZ Ceti  stars, since it opens the  interesting possibility of
probing  the  physical  mechanisms   operating  in  their  very  dense
interiors and,  particularly, to  obtain useful information  about the
crystallization  process which  occurs in  their cores  (Montgomery \&
Winget 1999).  This has been  motivated by the discovery of pulsations
in the  star BPM~37093 (Kanaan et  al.  1992), a massive  ZZ Ceti star
which has  a stellar mass of  $\sim 1.05 \, M_\odot$  and an effective
temperature $T_{\rm eff}\simeq  11800$~K, and which, therefore, should
have  a sizeable  crystallized  core  (Winget et  al.   1997).  It  is
interesting to note at this point  that this mass is very close to the
theoretical lower  limit ($\sim 1.05  \, M_\odot$) for  an oxygen/neon
white  dwarf to  be formed  (Salaris et  al.  1997;  Gil--Pons  et al.
2003).  Indeed, Gil--Pons et al.   (2003) find that for carbon burning
to take  place in the corresponding  progenitor, its mass  in the main
sequence should be $\sim 8.1 \,M_\odot$, leaving a massive oxygen/neon
white  dwarf of  $\sim 1.05  \, M_\odot$.  Hence, BPM~37093,  could be
either a carbon/oxygen or an oxygen/neon white dwarf, depending on the
precise value of its mass.  

We do  not intend to perform  a detailed modelling  of the pulsational
characteristics of  BPM~37093, since we  consider it to be  beyond the
scope of the paper. Instead, the aim of the present paper is to assess
the  adiabatic pulsational  properties  of massive  white dwarfs  with
carbon/oxygen  (CO) and oxygen/neon  (ONe) cores.   More specifically,
our  main goal  is to  explore the  possibility of  using  white dwarf
asteroseismology to distinguish between  both types of stars.  To this
end, we  compute the  cooling of massive  white dwarf models  for both
core  compositions taking  into  account the  evolutionary history  of
their progenitors and the  chemical evolution caused by time-dependent
element  diffusion during  the evolution.  The paper  is  organized as
follows.  In Sect.  2, we describe the main  characteristics of our CO
and  ONe  model  white  dwarfs.   Special emphasis  is  given  to  the
formation  of the  core chemical  profile during  the  pre-white dwarf
evolutionary stages. The pulsational predictions for the both types of
white dwarf  sequences are  described in  \S 3.  Finally,  in \S  4 we
briefly summarize our main findings and draw our conclusions.


\section{Input physics and evolutionary models}

In this work  we compute the evolution and  the pulsational properties
of massive white dwarfs with ONe cores. In pursuing this goal we adopt
the chemical profiles obtained by  Garc\'\i a--Berro et al. (1997) ---
see \S 2.2 for a detailed discussion.  Additionally, since a major aim
of our work  is to compare the theoretical  pulsational spectra of our
ONe white dwarf models with those of their CO analogs, we also compute
the evolution  of a white dwarf with  a CO core of  similar mass.  The
most  relevant results  of this  calculation are  presented in  \S 2.1
below.   An important  aspect  of  the present  study  common to  both
sequences   concerns   the  evolution   of   the  chemical   abundance
distribution  caused by diffusion  once the  white dwarfs  are formed.
This  is  particularly  relevant  for  the study  of  the  pulsational
properties of white dwarfs.  Indeed, element diffusion turns out to be
a key ingredient as far as mode trapping in pulsating stratified white
dwarfs is concerned (C\'orsico et  al.  2001; C\'orsico et al.  2002).
We  have  used a  time-dependent  treatment  for multicomponent  gases
(Burgers 1969) which accounts  for gravitational settling and chemical
and thermal  diffusion for $^{1}$H, $^{3}$He,  $^{4}$He, $^{12}$C, and
$^{16}$O species,  the dominant constituents  outside the core  of our
white  dwarfs  models ---  see  Althaus  et  al.  (2001a,  2001b)  for
additional  details.  In  this  way, the  trace element  approximation
usually invoked in most ZZ Ceti studies is avoided.

\subsection{Carbon/oxygen white dwarf models}

Our massive white dwarf models with  a CO core have been computed with
the evolutionary  code described at  length in Althaus et  al. (2003).
The code is based on  an up-to-date and detailed physical description.
Briefly,  the  code  uses   OPAL  radiative  opacities  for  arbitrary
metallicity  from Iglesias  \&  Rogers (1996)  and  from Alexander  \&
Ferguson (1994) for the low-temperature regime.  The equation of state
for the  low-density regime comprises partial  ionization for hydrogen
and helium  compositions, radiation pressure  and ionic contributions.
For  the  high-density  regime,  partially  degenerate  electrons  and
Coulomb interactions are also considered.  Under degenerate conditions
we  use an  updated  version of  the  equation of  state  of Magni  \&
Mazzitelli   (1979).   Neutrino   emission   rates  and   high-density
conductive   opacities  are  taken   from  the   works  of   Itoh  and
collaborators --- see Althaus et al.  (2002).  A nuclear network of 34
thermonuclear reaction  rates and 16  isotopes has been  considered to
describe hydrogen  (proton-proton chains and CNO  bi-cycle) and helium
burning.  Nuclear  reaction rates were  taken from Caughlan  \& Fowler
(1988) except  for the $^{12}$C($\alpha,\gamma)^{16}$O  reaction rate,
for which we adopted that of Angulo et al.  (1999).

\begin{figure}[t]
\centering
\includegraphics[clip,width=250pt]{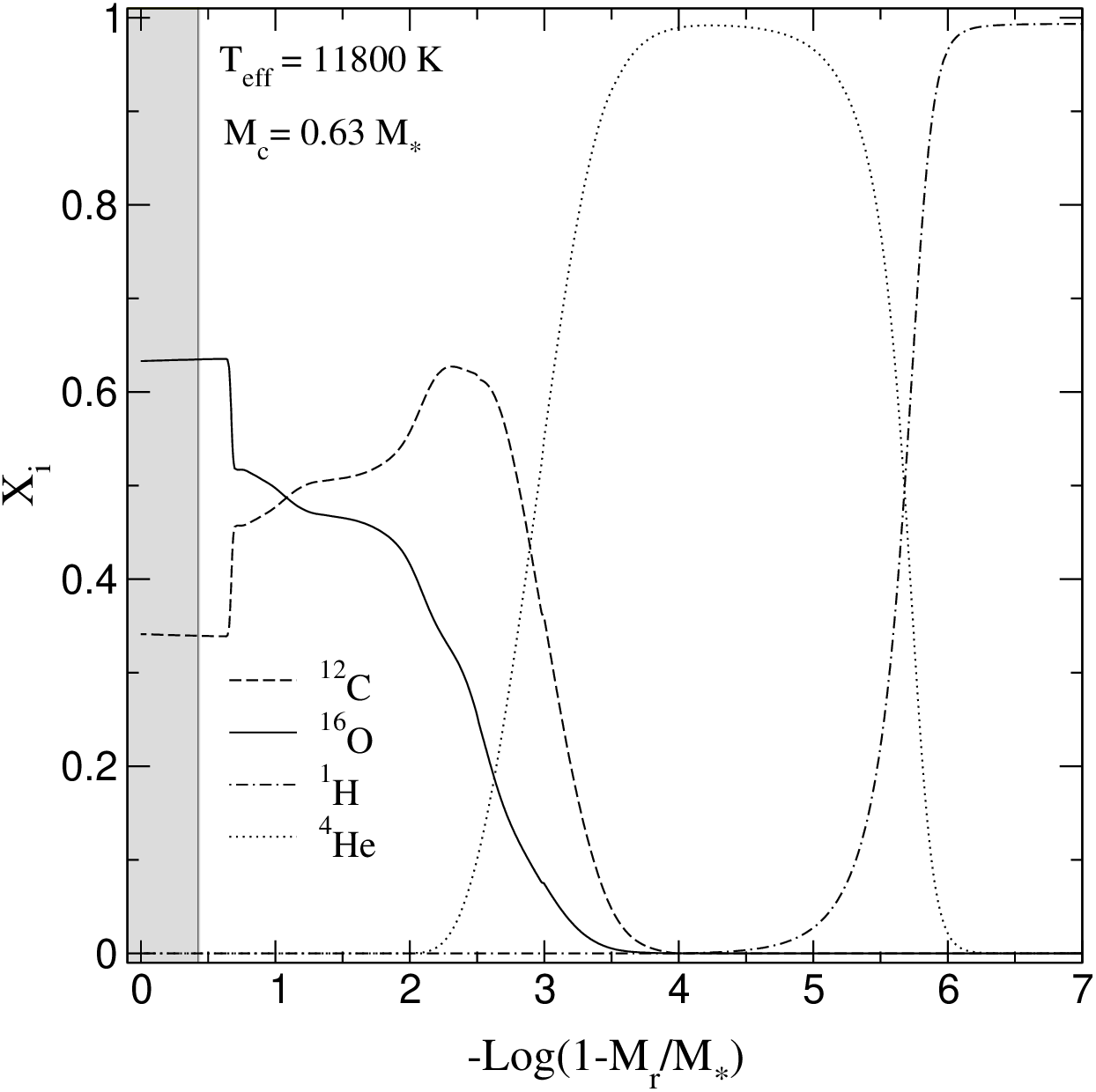}
\caption{The  internal  chemical profile  corresponding  to a $1.06 \, 
M_\odot$  CO  model  in  which chemical  rehomogenization  induced  by
crystallization has been neglected.  The gray area marks the domain of
crystallization.   $M_{\rm  c}$ means  the  crystallized  mass of  the
model.}
\label{quimi-co-s}
\end{figure}

The treatment  of the  change of the  abundances during  the pre-white
dwarf  evolution  is  an   important  aspect  to  be  considered.   In
particular, our code uses a time-dependent scheme for the simultaneous
treatment  of   chemical  changes   caused  by  nuclear   burning  and
convective, salt  finger and overshoot mixing, which  are described as
diffusion  processes  (Althaus  et  al.   2003).   We  have  used  the
Schwarzschild  criterion  for   convective  stability.   However,  the
occurrence  of additional  mixing  beyond what  is  predicted by  such
criterion is suggested by both theoretical and observational evidence.
In   particular,  extra   mixing  episodes   (particularly  mechanical
overshooting and/or  semiconvection) taking place  beyond the formally
convective boundary towards  the end of central helium  burning have a
large influence on  the carbon and oxygen distribution  in the core of
white dwarfs --- see Straniero  et al. (2003) for a recent discussion.
The occurrence of such mixing episodes leaves strong signatures on the
theoretical  period spectrum  of  massive ZZ  Ceti  stars (Althaus  et
al. 2003). Therefore, we have allowed for some mechanical overshooting
by following the formalism of Herwig (2000).  Our mixing scheme allows
for  a  self-consistent treatment  of  diffusive  overshooting in  the
presence  of  nuclear  burning.   In particular,  we  have  considered
exponentially  decaying  diffusive overshooting  above  and below  any
convective region,  including the  convective core (main  sequence and
central helium  burning phases), the external  convective envelope and
the short-lived helium-flash convection zone which develops during the
thermal  pulses.   Finally, convection  is  treated  according to  the
mixing length  theory for fluids with  composition gradients (Grossman
\& Taam 1996) that applies in the convective, semiconvective, and salt
finger instability regimes.  The mass  of the resulting CO white dwarf
is $\approx  0.94\, M_\odot$.   However, in order  to perform  a direct
comparison with the ONe white dwarf model previously mentioned we have
artificially  scaled the  stellar mass  of the  CO model  to  $1.06 \,
M_\odot$ --- approximately  the mass of BPM~37093 ---  but taking into
account the core chemical  distribution expected for a progenitor star
with a stellar mass correspondingly larger.

\begin{figure}
\centering
\includegraphics[clip,width=250pt]{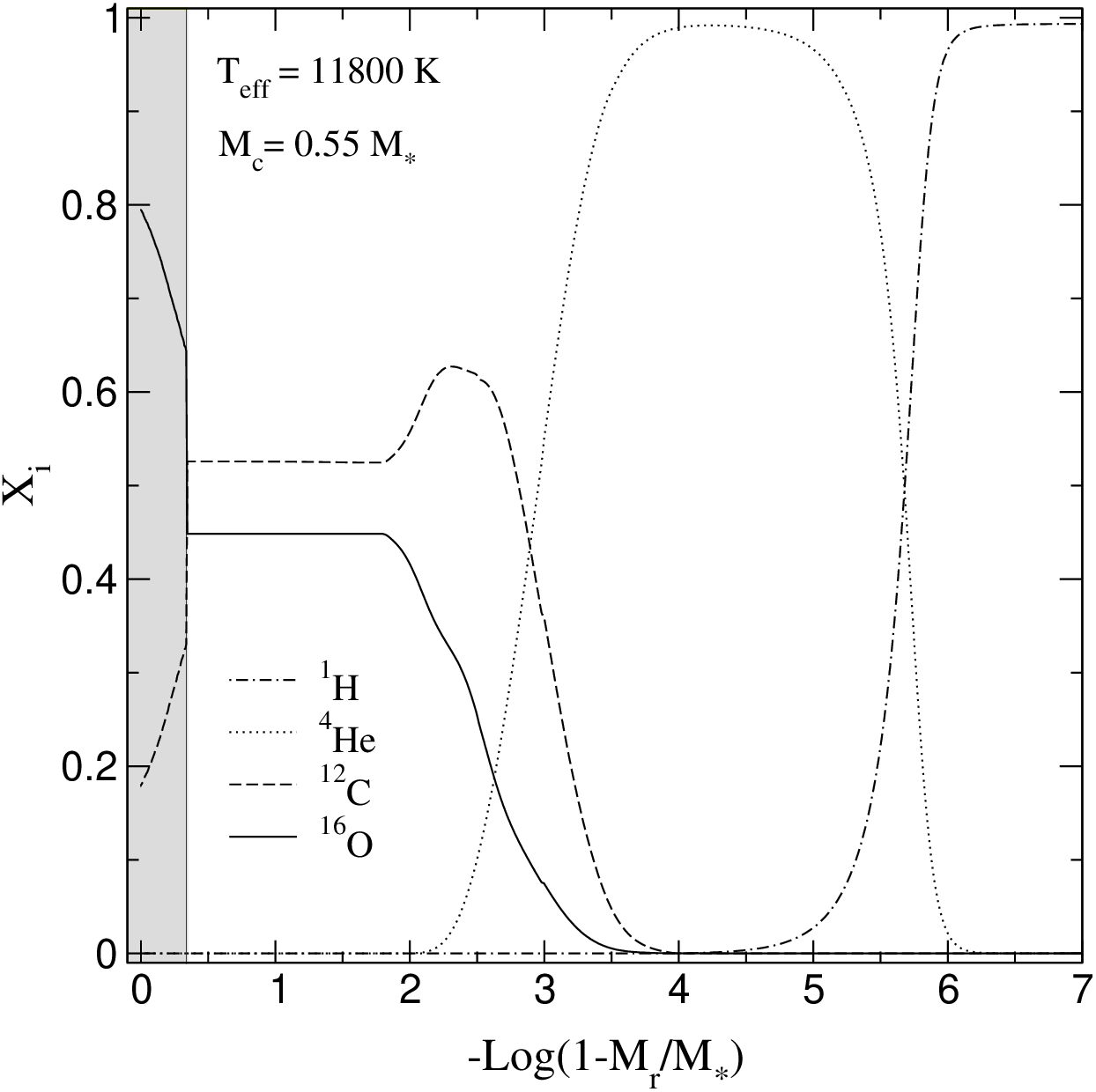}
\caption{Same  as  Fig.  \ref{quimi-co-s}, but  in the  case  in which 
chemical  rehomogenization   induced  by  crystallization   has  been
accounted for.}
\label{quimi-co}
\end{figure}

Crystallization in the  core of our ZZ Ceti models  and its effects on
their pulsational  pattern constitute an central point  of this paper.
Crystallization  sets in when  $\Gamma \equiv  Z^2 e^2  / \overline{r}
k_{\rm  B}  T=180$,   where  $\overline  r$  is  the   radius  of  the
Wigner-Seitz sphere.  In addition,  the chemical redistribution in the
fluid  above   the  solid  core  induced  by   phase  separation  upon
crystallization  (Garc\'\i a--Berro  et  al.  1988;  Segretain et  al.
1994) has been  taken into  account by adopting  the phase  diagram of
Segretain \& Chabrier (1993)  and following the procedure described in
Salaris  et al.   (1997)  ---  see, also,  Montgomery  et al.  (1999).
Nevertheless, for  the sake  of completeness we  have also  computed a
cooling sequence in which phase separation has been neglected.

In  our  evolutive  sequence,  the  CO white  dwarf  model  begins  to
crystallize at $T_{\rm eff} \approx 16600$ K. In Fig.~\ref{quimi-co-s}
we show the chemical profile of a  CO white dwarf model at the ZZ Ceti
stage, in  which the chemical  redistribution due to  phase separation
has been neglected.  The effective temperature of the model is $T_{\rm
eff} \approx 11800$ K and  their crystallized mass fraction amounts to
0.63  (the crystallized  core  is shown  as  a gray  zone).  Note  the
presence of a pronnounced step at $\log(1-M_{\rm r}/M_*) \approx -0.7$
in the oxygen profile.  This feature, which reflects the occurrence of
overshoot episodes prior  to the formation of the  white dwarf, leaves
strong  imprints on the  theoretical period  spectrum (Althaus  et al.
2003).  Note as well  that the external chemical interfaces, including
the He-core transition  are very smooth as consequence  of the element
diffusion processes acting during the evolution.

Fig.~\ref{quimi-co}  displays the chemical  profiles of  a CO  ZZ Ceti
model  in which chemical  redistribution due  to phase  separation has
been considered  at $T_{\rm eff} \approx  11800$ K.  In  this case the
crystallized mass  fraction amounts to 0.55, a  slightly smaller value
than  that  of  the  case  in which  phase  separation  was  neglected
(Fig.~\ref{quimi-co-s}).   This  is due  to  the  larger abundance  of
carbon  in  the  overlying  fluid  layers resulting  from  the  oxygen
enhancement in  the crystallized core of  the model.  Note  that, as a
consequence  of  the  mixing  resulting  from  phase  separation,  the
overshoot-induced  step  in  the  innermost oxygen  profile  has  been
completely wiped  out.  Consequently, the  theoretical period spectrum
for this model should be much less featured when compared with that of
the case in which  chemical rehomogenization has been neglected. This
feature will be discussed in depth in \S 3.

\begin{figure}[t]
\centering
\includegraphics[clip,width=250pt]{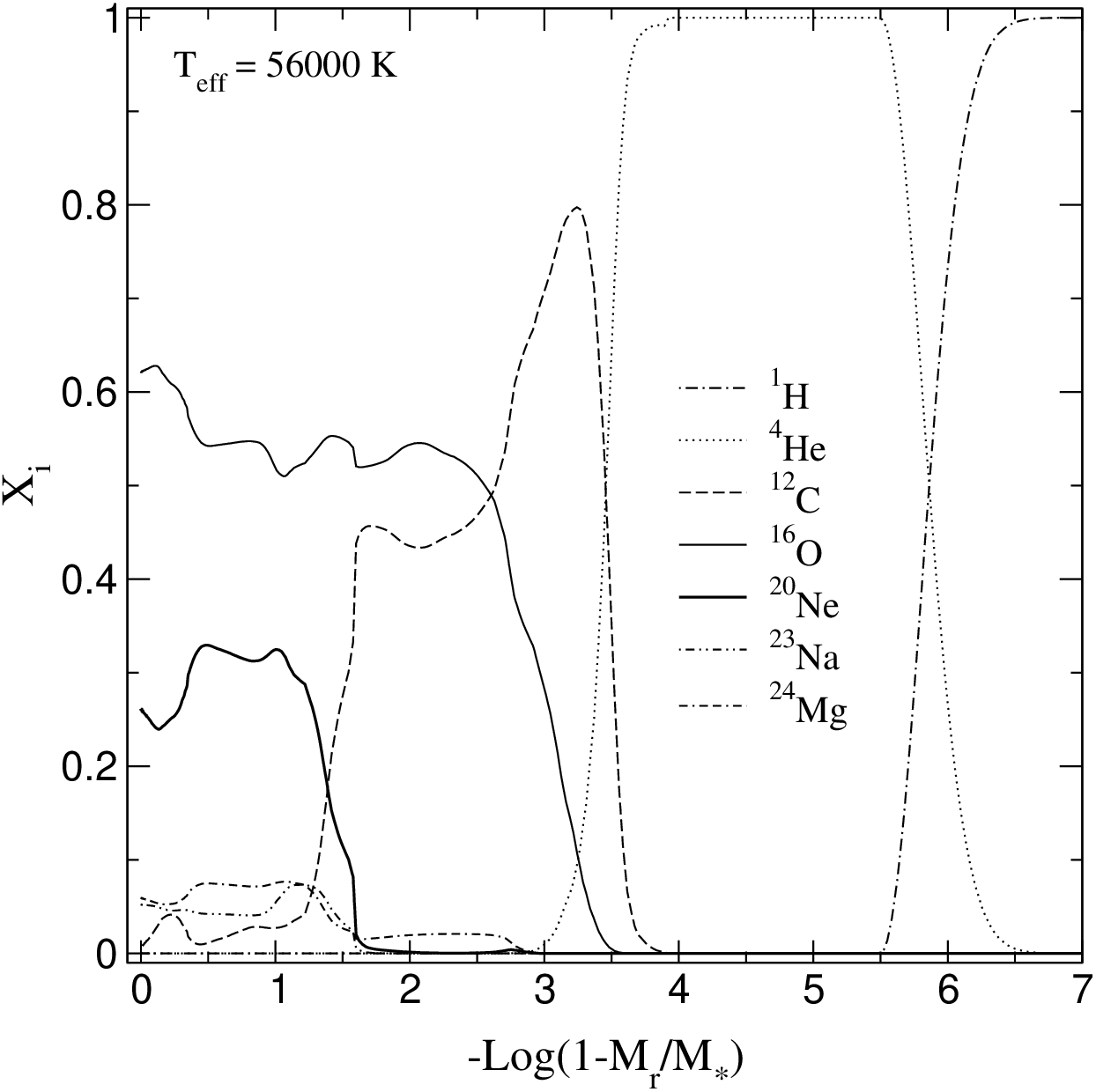}
\caption{The internal chemical profile corresponding to our initial 
ONe white dwarf model in terms of the outer mass fraction.}
\label{quimi-one-inicial}
\end{figure}

\subsection{Oxygen/neon white dwarf models}

The evolutionary  stages --- from the zero-age  main sequence, through
helium burning and off-center carbon ignition and burning in partially
degenerate conditions up  to the thermally pulsing AGB  --- leading to
the  formation of  ONe  white dwarfs,  as  well as  the input  physics
employed in the calculations are described in Garc\'{\i}a-Berro et al.
(1997) and  references  therein.    Here,  we  restrict  ourselves  to
summarize the  main characteristics  of the models  and, particularly,
their internal composition.  When  the carbon burning phase is already
finnished,  the  star is  left  with a  $\sim  1.06  \, M_\odot$  core
primarily composed  of $^{16}$O and  $^{20}$Ne, with trace  amounts of
$^{12}$C, $^{23}$Na  and $^{24}$Mg.  Surrounding  the core there  is a
buffer composed mainly  of a mixture of $^{12}$C  and $^{16}$O and, on
top  of it, the  overlying hydrogen-  and helium-rich  envelope layer.
The  initial envelope  chemical  profile ---  which, after  diffusion,
results   into    a   pure   hydrogen   layer   ---    is   shown   in
Fig.~\ref{quimi-one-inicial}.     We   have   obtained    an   initial
configuration  for our  ONe white  dwarf model  by simply  scaling the
internal  chemical  profiles   built-up  during  the  pre-white  dwarf
evolution to  the structure of the  massive, hot CO  white dwarf model
considered  previously.   In  this  way  we  obtain  a  good  starting
configuration  with  a  mass  close  to  that  of  BPM~37093  and  the
appropriate  chemical  profile.   Because  we are  interested  in  the
comparison of the pulsational  properties of highly-evolved CO and ONe
ZZ Ceti  models with the same  stellar mass, this  procedure is enough
for our purposes.

\begin{figure}
\centering
\includegraphics[clip,width=250pt]{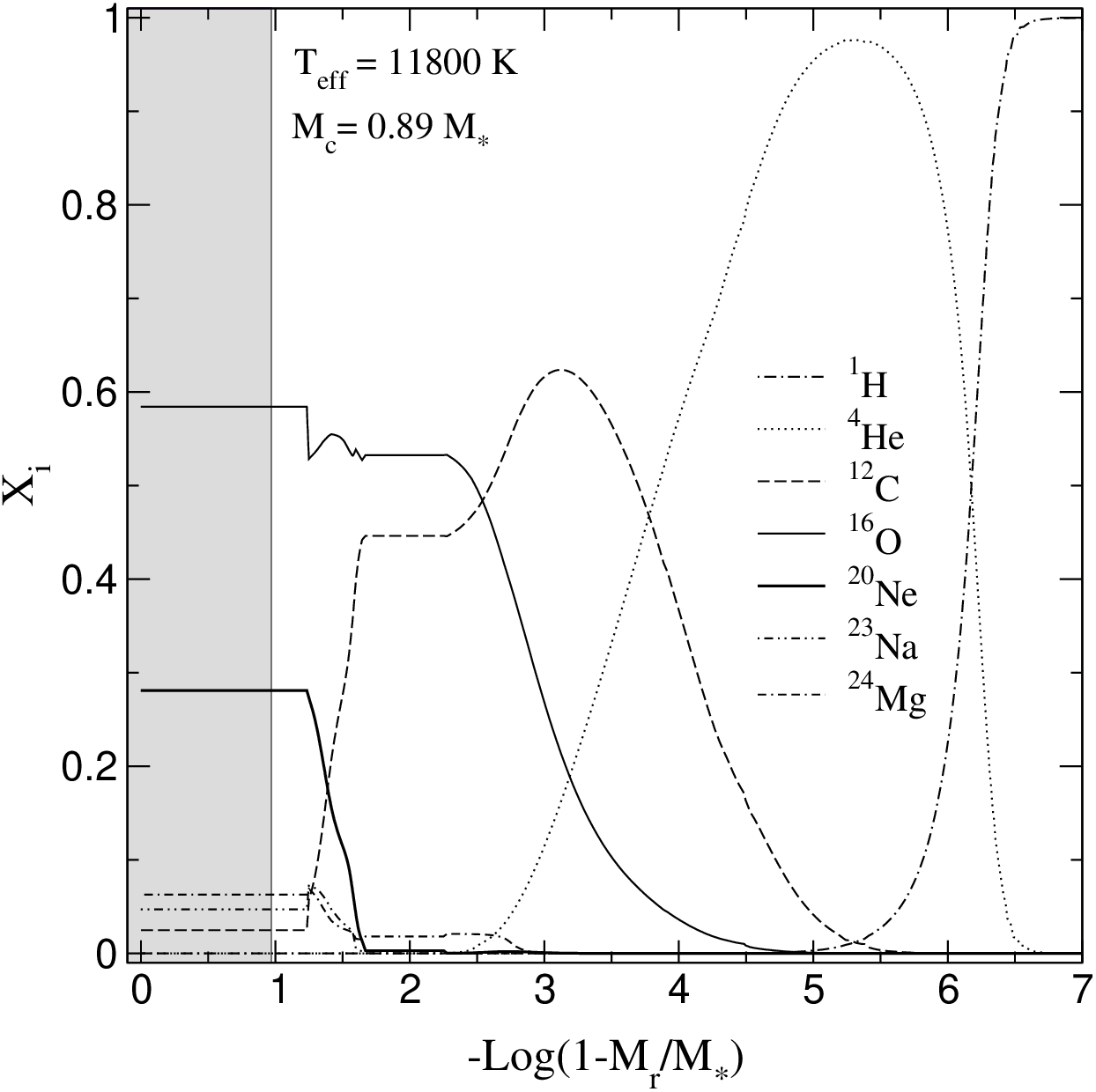}
\caption{Same as Fig.~\ref{quimi-one-inicial},  but for  an  ONe white 
dwarf model in the ZZ Ceti instability strip. The crystallized region
is displayed as a gray area.}
\label{quimi-one}
\end{figure}

In  our  evolutive sequence,  the  ONe  white  dwarf model  begins  to
crystallize at $T_{\rm eff} \approx 20700$ K, long before reaching the
ZZ Ceti  instability strip.  Fig.~\ref{quimi-one}  shows the resulting
chemical structure when the model  has reached the ZZ Ceti instability
strip. At this point of  the evolution, the percentage of crystallized
mass fraction amounts to $\simeq 90 \%$.  Two important points deserve
additional comments.   First, as it is  the case for  CO white dwarfs,
the effects  of chemical diffusion in the  external chemical abundance
distribution  is  clearly  noteworthy.   Indeed,  the  outer  chemical
interfaces  are  markedly  smoothed   out  by  diffusion.   Second,  a
rehomogenization process induced  by Rayleigh-Taylor instabilities has
led to a plateau in the  innermost chemical profile by the time the ZZ
Ceti stage  is reached. Note  that as a  result, a strong step  in the
chemical profile  at $\log(1-M_{\rm r}/M_*) \approx  -1.25$ has arisen
after rehomogenization,  a feature expected  to be the  most important
ingredient in determining  the structure of the period  pattern of the
model.  It is  as well important to mention here  that ONe white dwarf
models do not experience  a significant chemical redistribution due to
phase separation  because the  charge ratio between  Ne and O  is much
smaller  than that  of  O and  C  and, hence,  the  phase diagrams  of
Segretain \& Chabrier  (1993) do not predict a  sizeable enrichment in
Ne  in the solid  phase.  Moreover,  the evolutionary  calculations of
Ritossa et  al.  (1996), Garc\'\i a--Berro  et al. (1997)  and Iben et
al. (1997)  predict that after the  carbon burning phase  and when the
thermally pulsing  phase ensues there is a  non-negligible fraction of
unburnt carbon in  the oxygen/neon core. The mass  fraction of unburnt
carbon in  the ONe core can be  as high as $X_{\rm  C}\simeq 0.01$ ---
see  Fig.\ref{quimi-one-inicial}. The  same  happens for  the case  in
which a merger of two  light-weight white dwarfs is involved (Guerrero
et  al.  2004).    Hence,  the  calculation  of  the   effects  of  Ne
sedimentation  upon  crystallization  would  require a  ternary  phase
diagram (Segretain 1996) which, at present, is not yet well known and,
moreover, is beyond the scope of this paper.

\section{Pulsational calculations}

\begin{figure}
\centering
\includegraphics[clip,width=250pt]{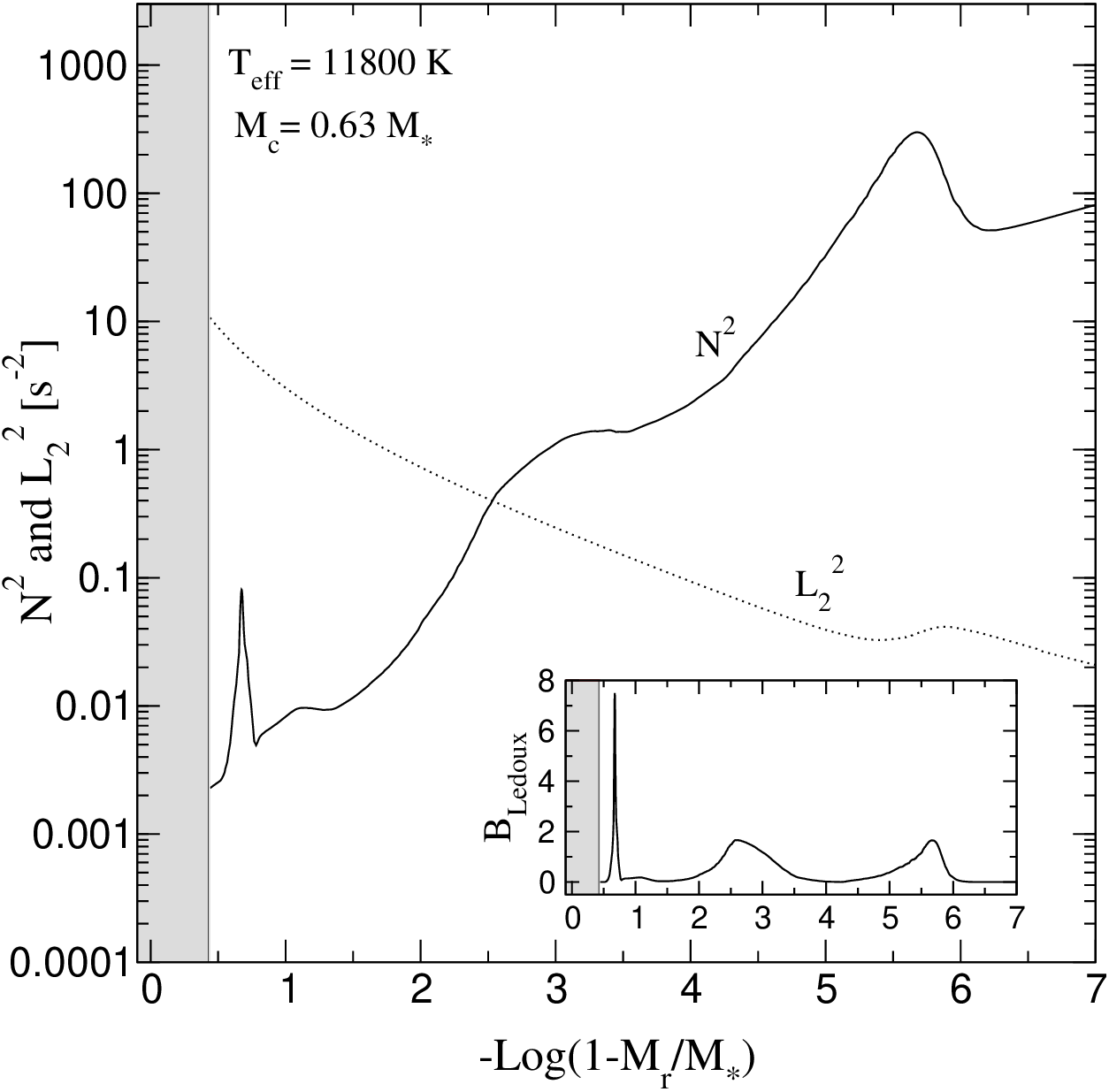}
\caption{The squared  Brunt-V\"ais\"al\"a frequency  ($N^2$) --- solid 
line ---  in terms  of the outer  mass fraction, corresponding  to the
same CO white dwarf  model shown in Fig.~\ref{quimi-co-s}.  Inset: the
Ledoux term $B$.   For the sake of completeness, we  also show using a
dotted line the acoustic  (Lamb) frequency corresponding to $\ell= 2$.
The gray area marks the domain of the crystallized core. Here, because
of  the   hard-sphere  approximation,   the  detailed  shape   of  the
Brunt-V\"ais\"al\"a frequency is irrelevant to pulsations and hence it
is not shown.}
\label{frec-co-s}
\end{figure}

We have  carried out  an adiabatic pulsational  analysis of  our white
dwarf models with the help of the same Newton-Raphson pulsational code
employed in C\'orsico et al.   (2001; 2002) and described in detail in
C\'orsico  (2003).  This  code  is coupled  to  the evolutionary  code
previously described.  However, for the purposes of this work, we have
done the appropriate modifications to the pulsational code in order to
properly  handle the  effects  of crystallization  on the  oscillation
eigenmodes.  Briefly,  the boundary  conditions at the  stellar center
(when crystallization has not yet  set in) and surface are those given
by Osaki \& Hansen (1973).  However,  when the core of the white dwarf
undergoes  crystallization  we  switch  the  fluid  internal  boundary
conditions  to  the  so-called  ``hard  sphere''  boundary  conditions
(Montgomery \& Winget 1999).   Within this approximation the nonradial
eigenfunctions are  inhibited to propagate in  the crystallized region
of the  core.  Following previous  studies of white  dwarf pulsations,
the  normalization condition  adopted  is $\xi_{\rm  r}/r=  1$ at  the
stellar  surface, being  $\xi_{\rm  r}$ the  radial  component of  the
displacement.   The  oscillation  kinetic  energy ($E_{\rm  kin}$)  is
computed according to  Eq.~1 of C\'orsico et al.   (2002), whereas the
weight function is computed as in Kawaler et al.  (1985).  Finally, we
derive  the asymptotic spacing  of periods  ($\Delta P_{\ell}$)  as in
Tassoul et al. (1990):

\begin{equation}
\Delta P_{\ell} = \frac{2 \pi^2}{\sqrt{\ell(\ell+1)}}\ 
\frac{1}{\int^{r_2}_{r_1} \frac{N}{r} dr}
\label{aps}
\end{equation}

\begin{figure}
\centering
\includegraphics[clip,width=250pt]{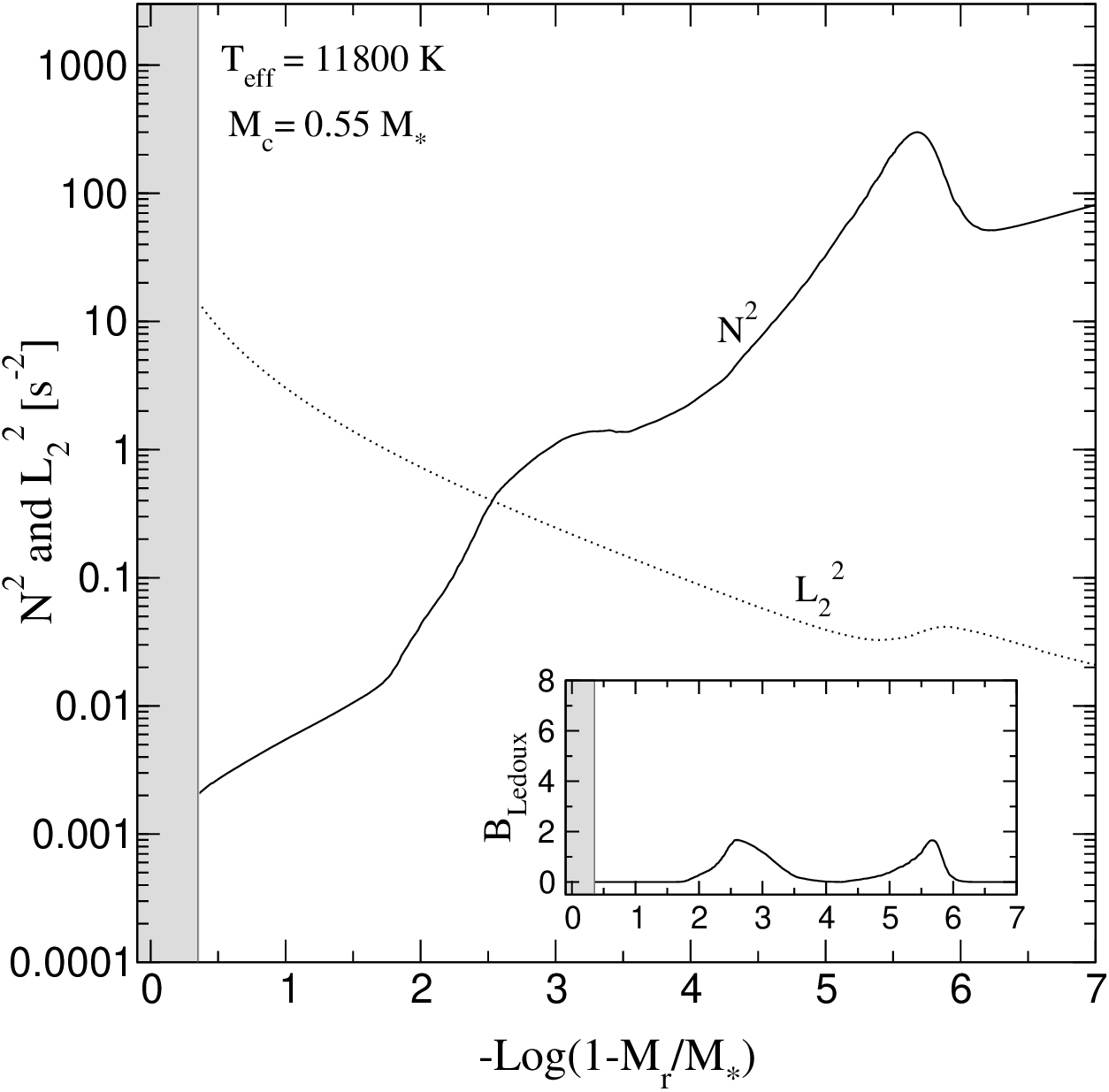}
\caption{Same as  in  Fig. \ref{frec-co-s}, but for the CO white dwarf 
model analyzed in Fig.~\ref{quimi-co}.}
\label{frec-co}
\end{figure}

\noindent where $N$  is the Brunt-V\"ais\"al\"a  frequency, and  $r_1$ 
and  $r_2$  are,  respectively,  the  radii of  the  inner  and  outer
boundaries of  the propagation region  of the modes.  Hence,  when the
core is already partially crystallized $r_1$ coincides with the radius
of the crystallization front,  $r_1=r(M_{\rm c})$.  It is important to
realize that the  turning point $r_1$ is no longer  a fixed value but,
instead, it is  a function of the crystallized  mass, $M_{\rm c}$ and,
hence,  of the  temperature of  the  (nearly) isothermal  core.  As  a
result, as the white dwarf  cools down, the internal boundary at $r_1$
moves outward, so the integral  $\int^{r_2}_{r_1} N dr / r$ decreases,
and,  consequently,  the  period  spacing increases  and  the  periods
themselves do so --- see Figs. 7 and 8 of Montgomery \& Winget (1999).

\subsection{Brunt-V\"ais\"al\"a frequency}

The  Brunt-V\"ais\"al\"a  frequency   is  computed  according  to  the
procedure  of  Brassard  et  al.  (1991).   This  numerical  treatment
accounts explicitly  for the  contribution to $N$  from any  change in
composition in the white dwarf models by means of the Ledoux term $B$.
This is an important aspect  in connection with the phenomenon of mode
trapping and mode  confining (Brassard et al.  1992;  C\'orsico et al.
2002). Specifically, the Brunt-V\"ais\"al\"a frequency is given by:

\begin{equation} \label{bvf}
N^2 = \frac{g^2\ \rho}{p}\ \frac{\chi_{_{\rm T}}}{\chi_{\rho}}\
\left(\nabla_{\rm ad} - \nabla + B \right). 
\end{equation}

\noindent The Ledoux term $B$, for the case of a multicomponent plasma,
is given by

\begin{equation}
B=-\frac{1}{\chi_{_{\rm T}}}\sum^{M-1}_{{\rm i}=1}
\chi_{_{X_{\rm i}}}
\frac{d\ln {X}_{\rm i}}{d\ln p}. 
\end{equation}

\noindent Here, $p$ is  the  pressure,  $\chi_{_{\rm T}}$  ($\chi_{\rm 
\rho}$)  denotes  the  partial logarithmic  derivative of the pressure  
with respect to  $T$ ($\rho$), $\nabla$ and $\nabla_{\rm  ad}$ are the
actual and adiabatic  temperature gradients, respectively, $X_{\rm i}$
is the abundance by mass of the chemical species $i$, and

\begin{equation} 
\chi_{_{X_{\rm i}}}= \left( \frac{\partial \ln{p}}
{\partial \ln{X_{\rm i}}} \right)_{\rho,T,\{X_{\rm j \neq i}\} }.
\end{equation}

We begin  by examining Fig.~\ref{frec-co-s},  where the square  of the
Brunt-V\"ais\"al\"a frequency corresponding to the same CO white dwarf
model  displayed  in  Fig.~\ref{quimi-co-s}  is shown.   It  is  worth
recalling that in this model phase separation upon crystallization was
disregarded. Note as  well that in our models, the  shape of the outer
chemical  interfaces  is   assessed  using  a  time-dependent  element
diffusion scheme. Consequently, and not surprisingly, our calculations
predict the presence  of very smooth bumps in the  profile of $N^2$ in
the external layers of the model (C\'orsico et al.  2002).  Hence, and
contrary  to what  it  is  usually found  when  the so-called  ``trace
element approximation''  is used,  in our calculations  these features
play a minor  role in the mode trapping  and confining characteristics
of the theoretical pulsational  spectrum.  Instead, it is the chemical
structure  of the  core what  determines the  structure of  the period
spectrum, as  already shown in Althaus et  al.  (2003).  Specifically,
the pulsation characteristics of this  model are fixed by the presence
of the pronounced peak in $N$ at $\log(1-M_{\rm r}/M_*) \approx -0.7$,
a  feature  directly related  to  the  overshoot-induced  step in  the
chemical profile (see Fig.~\ref{quimi-co-s}).
           
When phase  separation upon crystallization is taken  into account for
the CO  white dwarf models,  chemical redistribution in  the overlying
fluid  layers is  expected to  take place  as a  consequence of  the O
enhancement in the solid phase.  Consequently, the chemical abundances
(an their distribution throughout the fluid layer) of both C and O are
expected  to be  significantly different  from those  of  the previous
model.  The Brunt-V\"ais\"al\"a frequency  corresponding to a CO white
dwarf  model   in  which  the  chemical  profile   has  been  computed
self-consistenly  with the  predictions  of chemical  rehomogenization
(Fig.~\ref{quimi-co})  is shown in  Fig.~\ref{frec-co}.  Note  that at
this stage, the CO white dwarf models have a significant percentage of
its interior crystallized  ($55 \%$), in such a  way that the chemical
step  produced  by  core   overshooting  during  the  pre-white  dwarf
evolution has  been wiped out by the  ongoing chemical redistribution.
Because the external chemical transition regions have been smoothed by
diffusion, the shape of  the Brunt-V\"ais\"al\"a frequency has not any
abrupt feature  capable to strongly  pertubate the period  spectrum of
the model.  Hence,  the period spacing distribution should  be more or
less  uniform, at  least in  the asymptotic  limit of  large overtones
(long periods).

\begin{figure}
\centering
\includegraphics[clip,width=250pt]{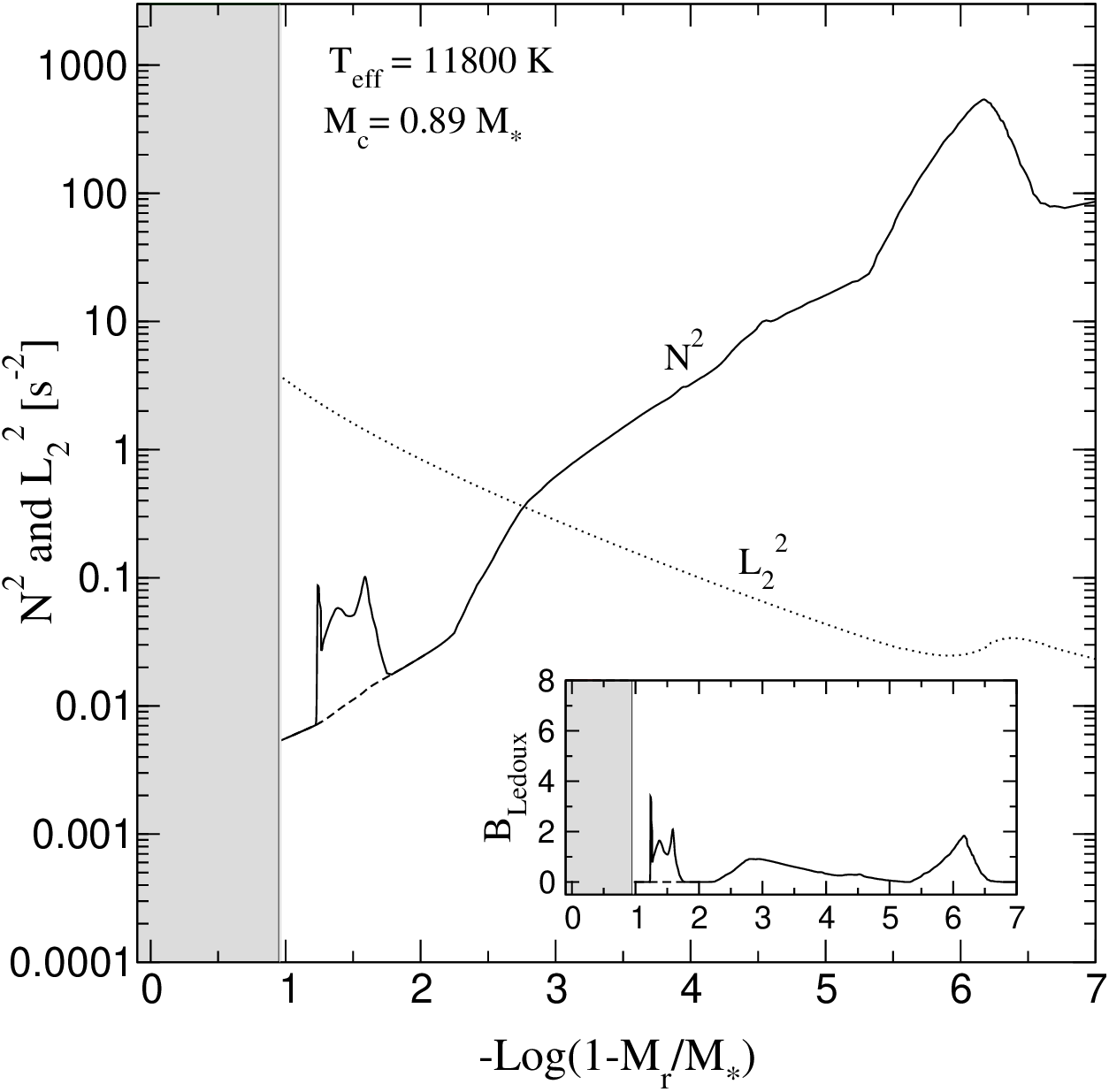}
\caption{Same as Fig.~\ref{frec-co}, but for the ONe white dwarf model
analyzed in Fig.~\ref{quimi-one}. Dashed  lines correspond to the case
in which  the Ledoux term $B$ is  neglected in the region $-2 \lesssim
\log (1-M_{\rm  r}/M_*)  \lesssim  -1$ (see  Sect. 3.2 for  additional 
details).}
\label{frec-one}
\end{figure}

Now,  we examine  the Brunt-V\"ais\"al\"a  frequency  characterizing a
typical   ONe  white   dwarf  model   at  the   ZZ  Ceti   stage.   In
Fig.~\ref{frec-one} we  plot the profile  of $N^2$ for the  same model
analyzed  in Fig.~\ref{quimi-one}.  Clearly,  the dominant  feature is
that located  at $-2 \lesssim \log(1-M_{\rm r}/M_*)  \lesssim -1$ (see
the Ledoux term  $B$ in the inset  of the figure). As we  shall see in
the   next   section,  this   feature   is   responsible  for   strong
non-uniformities in the period spectrum of this model.  

\subsection{Pulsational spectrum}

We have  computed adiabatic, nonradial,  spheroidal $g$(gravity)-modes
with $\ell= 2$ covering the  period range of pulsations observed in ZZ
Ceti  stars.  We  have  not considered  torsional  modes, since  these
modes,  characterized  by  very  short  periods  ---  of  up  to  20~s
(Montgomery  \& Winget 1999)  --- have  not been  detected in  ZZ Ceti
stars so far.

\begin{figure}
\centering
\includegraphics[clip,width=250pt]{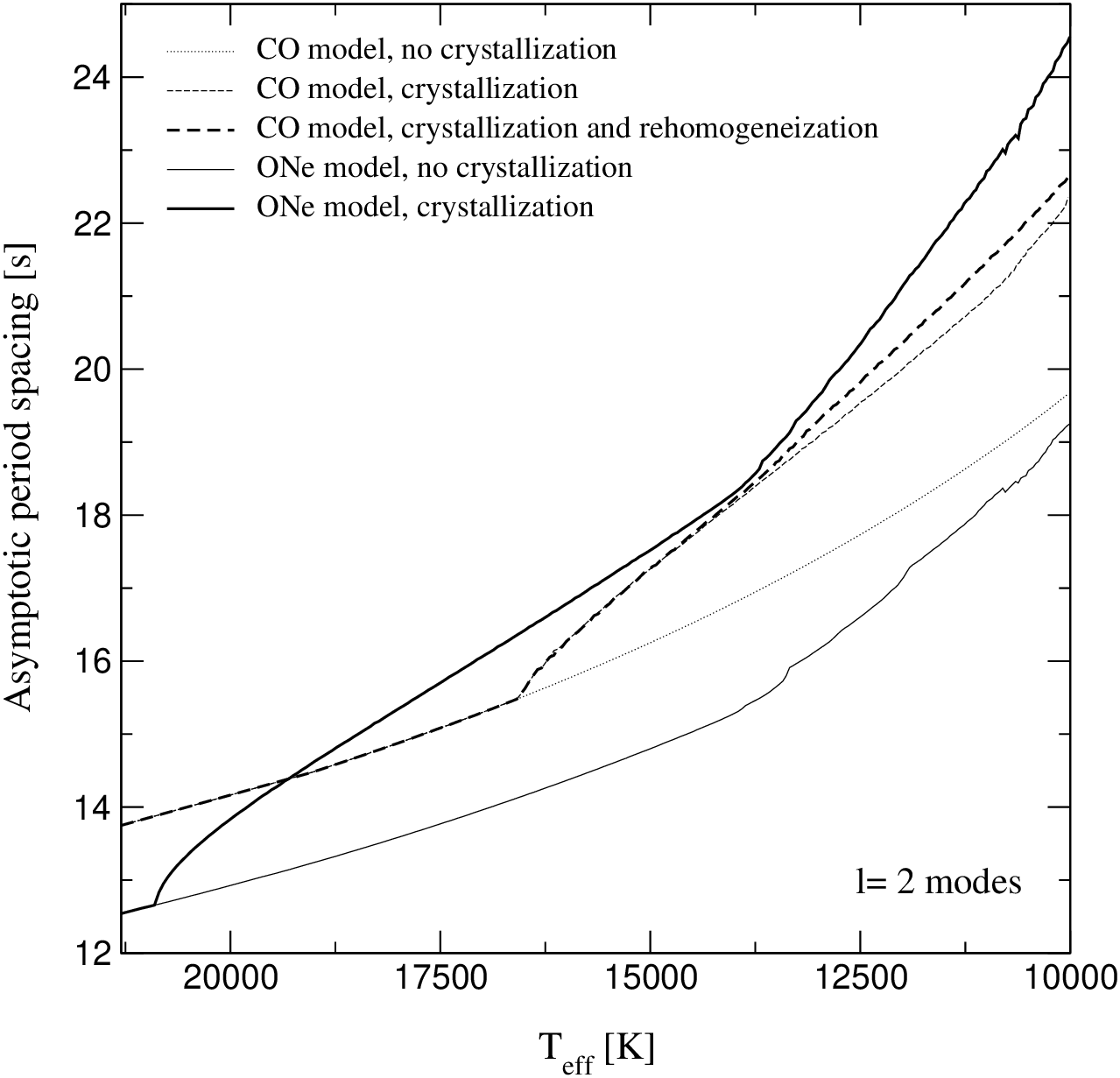}
\caption{Asymptotic   period   spacing,  $\Delta P_{(\ell=2)}$,  as  a 
function of the effective temperature, for the CO and ONe evolutionary
cooling sequences (thick and  thin curves, respectively). See text for
details.}
\label{pspacing}
\end{figure}

We begin by examining Fig.~\ref{pspacing}, where the asymptotic period
spacing as  a function of the  effective temperature is  shown for the
two CO  and the ONe white  dwarf models previously  discussed, and for
two  additional models  in which  crystallization has  been completely
disregarded.  These  two additional models  are only computed  for the
sake  of  comparison  and  in   order  to  exemplify  the  effects  of
crystallization  in spite  of the  fact that  they are  not physically
sound.  The thick  lines correspond to our fiducial  models.  That is,
the  thick   solid  line  corresponds   to  an  ONe  model   in  which
crystallization has been taken  into account, whereas the thick dashed
line  corresponds  to   the  CO  white  dwarf  model   in  which  both
crystallization  and phase  separation were  taken into  account.  The
thin dashed  line corresponds to  a CO model in  which crystallization
but no phase  separation was considered. On its  hand, the thin dotted
line corresponds  to a CO  model in which neither  crystallization nor
phase  separation were accounted  for.  Finally,  the thin  solid line
indicates the  results for an  ONe model in which  crystallization was
completely disregarded.  The  results presented in Fig.~\ref{pspacing}
deserve  several  comments.  Firstly,  for  all  the cases  considered
$\Delta  P_{\ell=2}$ is an  increasing function  of $T_{\rm  eff}$, as
expected  from   the  fact  that   the  Brunt-V\"ais\"al\"a  frequency
decreases  as  the  white  dwarf  cools  down  ---  $\chi_{_{\rm  T}}$
decreases as  degeneracy increases, see  Eq.~(\ref{bvf}). Secondly, it
is noteworthy that  in the case in which  crystallization is neglected
$\Delta  P_{\ell=2}$ is smaller  for the  ONe models  than for  the CO
ones.   This trend  is understood  in  terms of  $N^2$ having  notably
larger values at the core region  in the ONe models compared to the CO
models.  Thus, the integral of  Eq.~(\ref{aps}) turns out to be larger
in the  case of  ONe models and,  consequently, the  asymptotic period
spacing   is  smaller.    

In  contrast,  for  the   cases  in  which  crystallization  has  been
considered, the  values of $\Delta  P_{\ell=2}$ of the ONe  models are
always  larger than  those  of the  CO  ones.  In  part,  this can  be
understood on the fact that the ONe model is characterized by a larger
crystallized  mass  fraction than  the  CO  one  for a  fixed  $T_{\rm
eff}$.  This   in  turn  implies   that  the  integral  of   $N/r$  in
Eq.~(\ref{aps}) is smaller and the resulting asymptotic period spacing
in the  ONe model is  larger.  With regard  to the CO models,  we note
that,  when  chemical rehomogenization  is  considered the  asymptotic
period spacing is slightly greater when compared with the situation in
which  rehomogenization is neglected.  This can  be understood  on the
basis that,  in the case in  which rehomogenization is  not allowed to
operate,  the integral  in Eq.~(\ref{aps})  has an  extra contribution
from the overshoot-induced step  via the strong peak at $\log(1-M_{\rm
r}/M_*)\approx  -0.7$ in  the  Brunt-V\"ais\"al\"a  frequency (see
Fig.~\ref{quimi-co-s}).

\begin{figure*}[htb]
\centering
\includegraphics[clip,width=500pt]{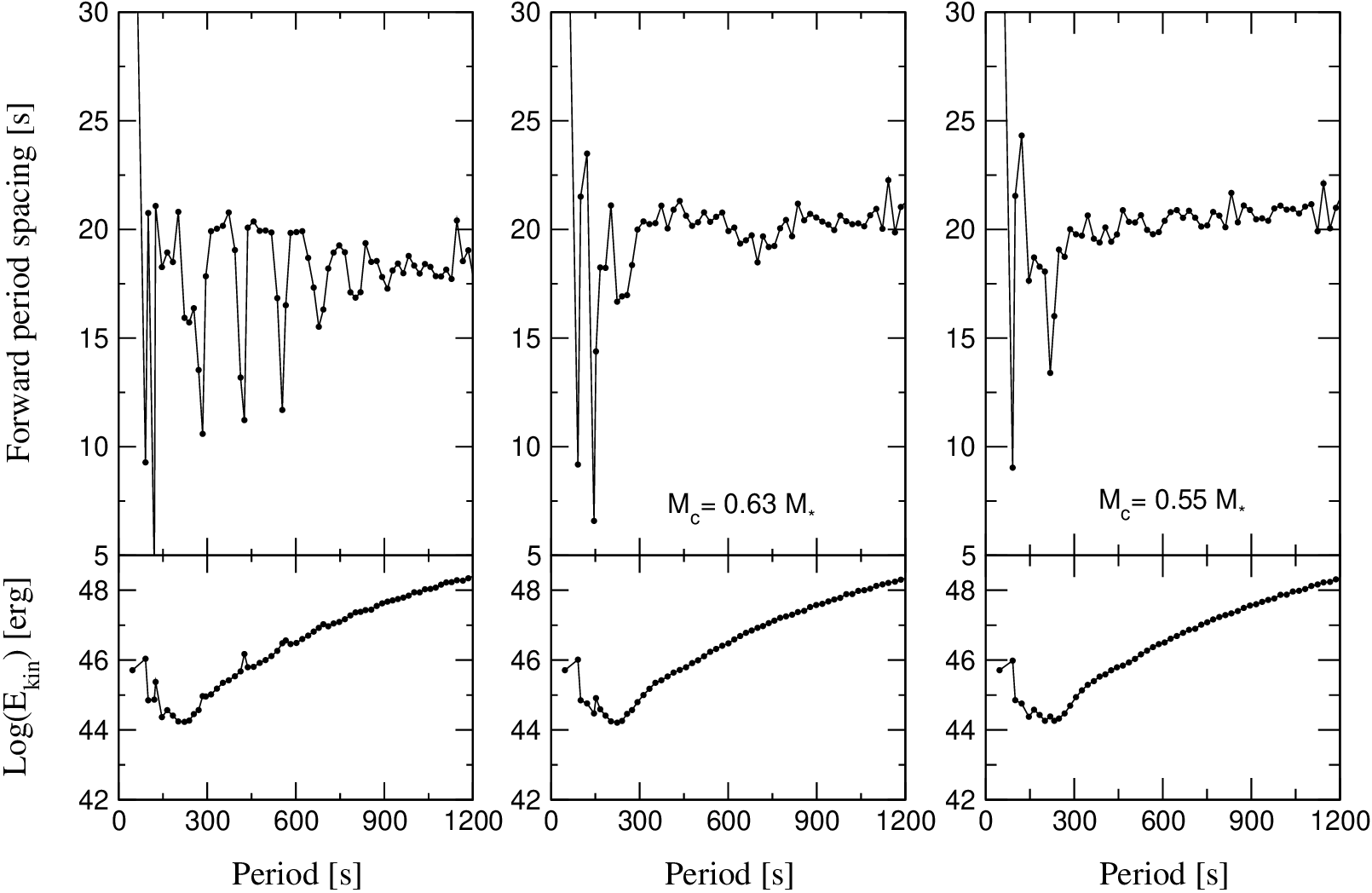}
\caption{The  forward  period spacing  (upper panels)  and the kinetic 
energy (lower panels)  in terms of the periods  of $\ell= 2$ pulsation
modes.  The left panels correspond to  a CO white dwarf model in which
crystallization has  not been taken  into account. The  central panels
show the  same quantities  for a  model at the  same $T_{\rm  eff}$ in
which   crystallization   but    no   phase   separation   have   been
considered.  Finally, the  right panels  show the  case in  which both
crystallization and phase separation have been accounted for.}
\label{dp_ek_COs}
\end{figure*}

We now  examine the  forward period spacing,  defined as  $\Delta P_k=
P_{k+1}-P_k$ (being $k$ the  radial overtone). This quantity, which is
particularly sensitive  to the details of  internal chemical structure
of  the white  dwarf models,  is usually  employed to  infer  the mode
trapping properties  of ZZ Ceti  stars.  In addition,  the theoretical
period spacing  can be  directly compared with  the observed one  in a
given star, whenever a  sufficient number of consecutive periods (with
the same  harmonic degree  $\ell$) can be  measured. Here  we restrict
ourselves  to present  pulsational results  corresponding to  the same
models   analyzed  in   Figs.    \ref{frec-co-s},  \ref{frec-co}   and
\ref{frec-one}. We  recall that these  models are characterized  by an
effective temperature of $\approx 11800$ K, which is representative of
ZZ Ceti stars. Fig.  \ref{dp_ek_COs} shows the $\ell=2$ period spacing
distribution  (top  panels)  and  the kinetic  energy  of  oscillation
(bottom panels)  as a function of the period  for a CO white  dwarf in
which crystallization has  been ignored (left panels), for  a CO white
dwarf in  which crystallization  was taken into  account but  no phase
separation was  considered (middle panels)  and for the case  in which
both crystallization and phase  separation were properly accounted for
(right panels).

Clearly,  when  crystallization   is  considered  the  period  spacing
distribution  is much  less  featured.  A  similar  behavior has  been
reported by Montgomery  \& Winget (1999) in the  frame of calculations
in  which the  crystallized  mass  fraction is  considered  as a  free
parameter.   When crystallization  is ignored,  the  internal boundary
condition  remains fixed  at the  center of  the star.   The resulting
eigenspectrum  is composed  by  some core-confined  modes, some  modes
trapped in the outer layers, and the remainder normal modes (C\'orsico
et al.   2002; Althaus et al.   2003).  In this case  the modes nearly
preserve  their character  during  the evolution.   In contrast,  when
crystallization  is  considered  (middle  panels) the  inner  boundary
condition  moves  together with  the  crystallization  front, and  the
eigenfunctions  of all  modes are  pulled out  to the  surface  of the
model.    Under   these   circumstances,  the   resonance   conditions
determining which modes are  trapped or confined are strongly modified
as the degree of crystallization  increases.  This is clearly shown in
the top  middle panel of Fig.~\ref{dp_ek_COs}, which  corresponds to a
model in  which the  crystallization front is  located at  $M_{\rm c}=
0.63 \, M_*$.  Note that for the high- and intermediate-overtone modes
the strong minima  (associated with core-trapped modes) characterizing
the $\Delta  P_k$ distribution of the model  for which crystallization
was  ignored  (left panel),  are  not present  in  the  case in  which
crystallization  is considered (middle  panel).  A  look to  the lower
panels of Fig.~\ref{dp_ek_COs} shows as well that when crystallization
is taken into account the  differences between the kinetic energies of
trapped, normal  and confined modes are  smaller.  Another interesting
point is  that the mean period spacing  increases when crystallization
is  considered, in agreement  with the  predictions of  the asymptotic
theory  of pulsations  and with  the results  of Montgomery  \& Winget
(1999).

We now  examine the more realistic  case in which  phase separation is
included in  addition to crystallization  (right panels).  In  view of
the  above   considerations,  we   expect  a  smooth   period  spacing
distribution.   We note  that the  period spacing  distribution (upper
right  panel) has  a fairly  clean structure  for periods  longer than
$\approx 300$ s.    In fact, the forward period spacing for modes with
$k> 12$  becomes very close  to $20.6$ s,  the value predicted  by the
asymptotic  theory, given by  Eq.~(\ref{aps}).  As  already discussed,
this is due mostly to the absence of any abrupt feature in the profile
of  the   Brunt-V\"ais\"al\"a  frequency.   In  fact,   the  model  is
characterized by a crystallized mass fraction of $\approx 0.55$, which
is large  enough for the ongoing  chemical rehomogenization processes
to  erase the  overshoot-induced  step in  the  chemical profile  (see
Figs.~\ref{quimi-co-s} and \ref{quimi-co}).

\begin{figure*}
\centering
\includegraphics[clip,width=350pt]{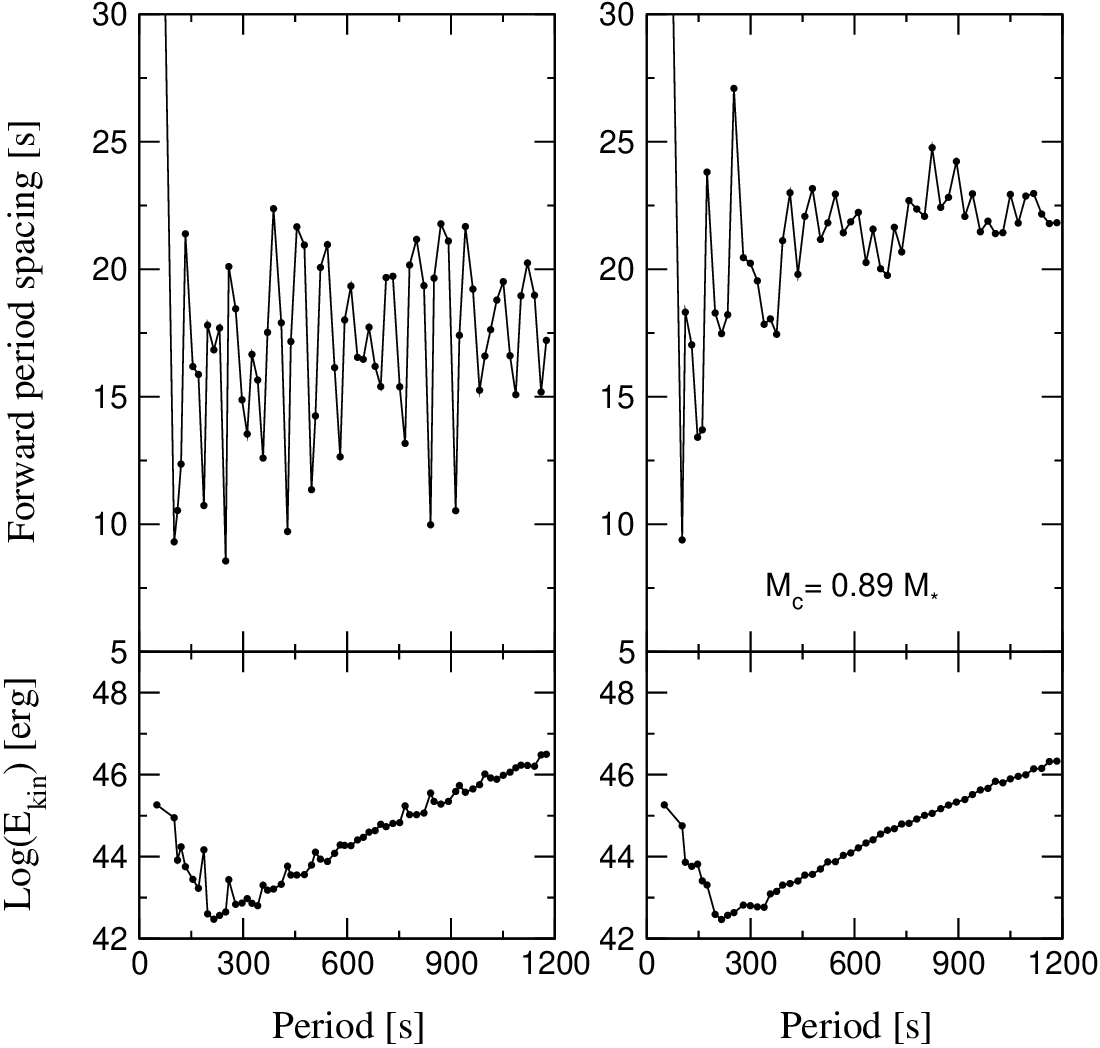}
\caption{Same as  Fig.~\ref{dp_ek_COs}, but  for the  case  of an  ONe 
white dwarf.} 
\label{dp_ek_ONE}
\end{figure*}

As previously  stated, the major  aim of this  work is to  explore the
possibility  of  using  white  dwarf asteroseismology  to  distinguish
massive CO  white dwarfs  from those having  ONe cores.  In  the right
panels of  Fig.~\ref{dp_ek_ONE} we  show the distributions  of $\Delta
P_k$  and $E_{\rm  kin}$ corresponding  to our  fiducial ONe  model at
$T_{\rm  eff} \approx 11800$  K (which  corresponds to  a crystallized
mass $M_{\rm c} \approx 0.90\, M_*$).  For the sake of completeness in
the left panels of this figure we also show the same distributions for
the  case  in  which  crystallization  was  ignored.   We  note  that,
irrespective  of the  assumptions regarding  crystallization  for both
types of models (CO or ONe), the period spacing diagrams of ONe models
are notably different from those of the CO models.  In particular, for
the case  in which crystallization was disregarded  (left panels), the
period  spacing diagram  of the  ONe model  shows strong  and abundant
features linked with mode trapping and confining, mostly driven by the
structure at  $-2 \lesssim \log(1-M_{\rm  r}/M_*) \lesssim -1$  in the
Brunt-V\"ais\"al\"a frequency  profile (see Fig.~\ref{frec-one}). When
crystallization is taken into account, the period spacing distribution
of the  ONe model  still has  a complex structure  but with  a notably
lower  strength  (right  upper  panel  of  Fig.~\ref{dp_ek_ONE}).   In
particular, we  note the  presence of three  minima at  $\approx 300$,
$600$ and  $900$ sec.  However,  the comparison of the  period spacing
diagrams   in   the   right   panels  of   Figs.~\ref{dp_ek_COs}   and
\ref{dp_ek_ONE} indicates  that two models with the  same stellar mass
and effective  temperature but differing in the  core composition (one
having a core rich in carbon and oxygen, and the other mainly composed
by  oxygen and  neon) should  be  characterized by  a quite  different
pulsational spectrum.

An additional  difference in  the pulsational properties  between both
types of  models is found when  the kinetic energies of  the modes are
compared. In  fact, {\sl all}  the modes of  the ONe model  have lower
energies than those of their  counterparts of the CO model.  Two facts
help  in clarifying  this  issue.   We recall  that  $E_{\rm kin}$  is
proportional to the  density and to the squared  radial and horizontal
eigenfunctions.   Firstly,   the  propagation  region   in  which  the
eigenmodes   are   allowed   to   oscillate  have   different   sizes.
Specifically, for the ONe white  dwarf at $T_{\rm eff } \approx 11800$
K,  we   find  that   $90  \%$  of   the  stellar  mass   has  already
crystallized. Hence,  in this high-density  region nonradial $g$-modes
are excluded.  Therefore,  the modes are capable to  propagate only in
the remaining  external, low-density  region.  In the  case of  the CO
model at  the same  $T_{\rm eff}$, the  crystallized mass  fraction is
aproximately  0.55,  and  so   the  modes  can  propagate  in  regions
characterized by considerably high densities.  Secondly, we have found
that for the  case of ONe models the  amplitudes of the eigenfunctions
are considerably  smaller than those  of their counterparts in  the CO
models.  As  a final remark, we  note that the mean  period spacing of
the  pulsation modes  corresponding to  ONe white  dwarfs  is slightly
larger than  that of the CO  models.  This is again  in agreement with
the   predictions   of   the    asymptotic   theory,   as   shown   in
Fig.~\ref{pspacing}.

\begin{figure}
\centering
\includegraphics[clip,width=250pt]{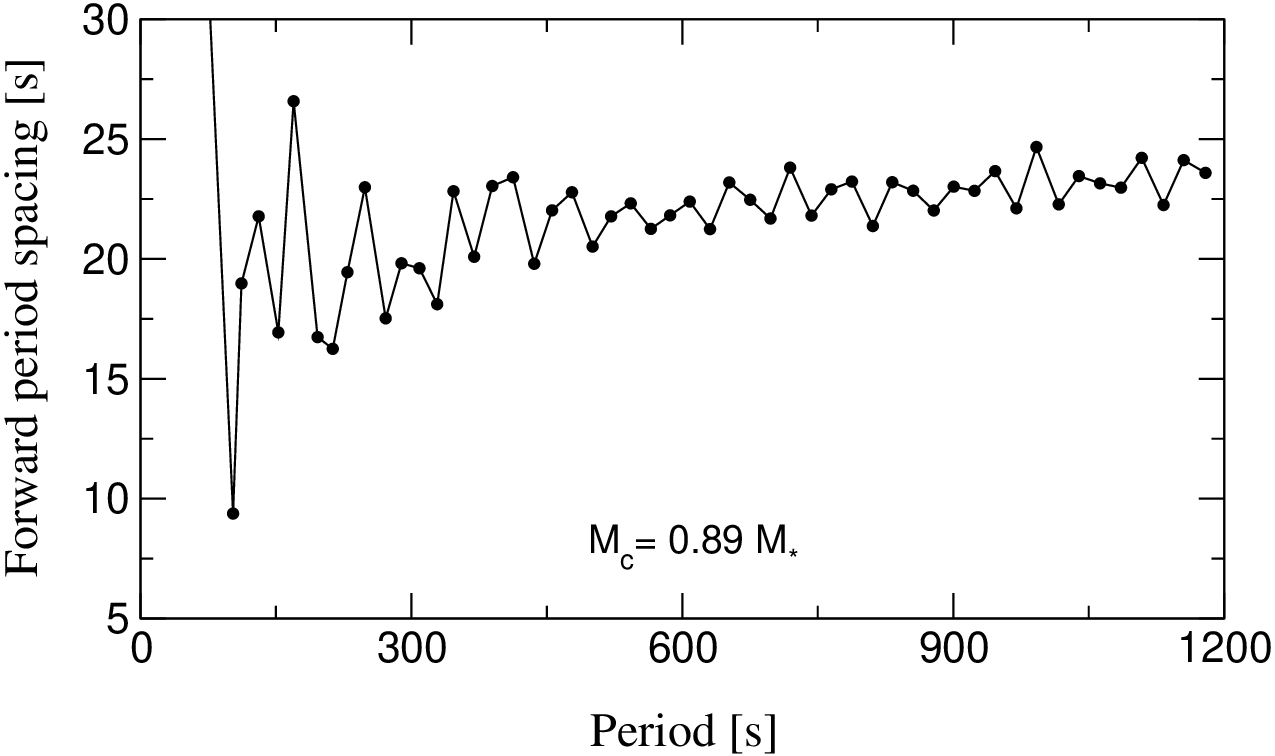}
\caption{Same as the upper-right panel of Fig. \ref{dp_ek_ONE}, but for 
the  case in  which the  effect  of the  strong step  in the  chemical
profile  at  $-2  \lesssim  \log(1-M_{\rm r}/M_*)  \lesssim  -1$  (see
Fig.  \ref{quimi-one})   on  the  shape   of  the  Brunt-V\"ais\"al\"a
frequency has  been neglected  by setting $B=  0$ in that  region (see
Fig. \ref{frec-one}).} 
\label{dp_ONE_bcero}
\end{figure}

In closing,  we stress  that the main  differences in  the pulsational
spectrum between ONe and CO models are due to the presence of a strong
step in the core chemical profile of the ONe model. As mentioned, this
step marks the  outer edge of the mixing zone  induced by the negative
molecular weight in the deeper layers of the ONe  white dwarf progenitor. 
However, it could
be  possible  that additional  mixing  episodes  beyond this  external
border may wipe  out the chemical step. This being  the case, we would
expect  the   period  pattern  to   be  modified  to   a  considerable
extent. This is indeed borne out by Fig.  \ref{dp_ONE_bcero}, in which
the period spacing is computed under the assumption that $B= 0$ in the
region  at $-2  \lesssim \log(1-M_{\rm  r}/M_*) \lesssim  -1$  and the
resulting run  of the Brunt-V\"ais\"al\"a frequency  is that displayed
with  dashed line  in  Fig.  \ref{frec-one}. We  are  aware that  this
procedure  does not  eliminate completely  the effects  of  the strong
chemical feature on the period spectrum, because the change in density
associated  with  this chemical  interface  is  still  present in  the
computation   of  $N$  (see   Eq.  2).   However,  we   consider  this
approximation  reliable enough for  our purposes  here. Note  that the
forward period spacing becomes much less featured as compared with the
case   in  which   the  Brunt-V\"ais\"al\"a   frequency   is  computed
self-consistently (right panel  of Fig. \ref{dp_ek_ONE}), and instead,
it is reminiscent of the results  obtained for the CO model (see right
panel of Fig.  \ref{dp_ek_COs}).  This aspect would render the current
asteroseismological techniques  much less capable to  infer the actual
core composition of the massive white dwarfs. We note, however,  that
the average period spacing remains virtually unchanged even in presence
of such extra mixing episodes.

\section{Conclusions} 

In this  work we have computed  the evolution of  massive white dwarfs
with hydrogen-rich  envelopes and carbon/oxygen  and oxygen/neon cores
with  the   major  aim   of  comparing  their   adiabatic  pulsational
properties.  Particular  attention was given  to the formation  of the
core chemical  profile during the pre-white  dwarf evolutionary stages
and to the evolution of  the chemical abundance distribution caused by
element  diffusion during the  white dwarf  regime.  In  addition, the
chemical   rehomogenization   induced   by   phase   separation   upon
crystallization  as well  as  the effects  of  crystallization on  the
oscillation eigenmodes have been fully taken into account.

Our results show that because  the chemical interfaces in the envelope
are markedly smoothed by diffusion, the chemical structure of the core
mostly determines  the structure of the period  spectrum.  This result
reinforces  the conclusions  arrived at  in C\'orsico  et  al. (2001),
(2002) and  Althaus et  al. (2003)  about the  role  of time-dependent
element diffusion in white dwarf pulsation calculations.  Moreover, we
find  that, in  contrast with  their carbon/oxygen  core counterparts,
oxygen/neon white dwarfs  are characterized by strong non-uniformities
in their period  spacing distribution.  The lack of  a featured period
spacing  distribution characterizing  white dwarfs  with carbon/oxygen
cores arises  from the fact that  the mixing episode  induced by phase
separation  as  crystallization  proceeds  completely  wipes  out  any
feature in the shape of the innermost chemical profile built up during
the  evolutionary  stages prior  to  the  white  dwarf formation.   In
addition, we found that  the average period spacing characterizing our
oxygen/neon   models  is   appreciably   larger  than   that  of   the
carbon/oxygen counterparts having the  same stellar mass and effective
temperature. This is true even in the presence of extra mixing episodes 
that could eventually remove the sharp abundance discontinuity left in the oxyen n/neon models. Finally,  we also find  a quite different  kinetic energy
spectrum for both types of models.

Hence, it  turns out  that the asteroseismological  techniques usually
employed in variable white dwarf studies are, in principle, a powerful
tool for distinguishing massive  carbon/oxygen white dwarfs from those
having oxygen/neon  cores. In  this regard, the  most massive  ZZ Ceti
star presently  known, BPM~37093, is of particular  interest since the
value of its stellar mass of $\sim 1.05 \, M_\odot$ places it close to
the  theoretical lower  limit  for  a oxygen/neon  white  dwarf to  be
formed. Nonetheless, the scarcity of detected  periods (eight periods) 
in the light curve of BPM~37093 renders conclusive asterosesimological 
inferences in this case very difficult. However, since the average period 
spacing is larger for the oxygen/neon models than for the carbon/oxygen 
ones, then a comparison with the observed value 
would help to get a clearer insight into the actual core composition of 
this massive white dwarf. We feel that this aspect would deserve to be 
explored in the frame of a set of white dwarf models with various stellar 
masses. Ongoing  attempts  to  constrain 
the fundamental properties of BPM~37093 using asteroseismological data
would  allow us  to shed  new light  on the  scenarios leading  to the
formation of  single massive white dwarfs  and on the  nature of their
progenitors.

\begin{acknowledgements}
This work was partially supported by the Instituto de Astrof\'{\i}sica
La Plata, by the MCYT  grants AYA04094--C03-01 and 02, by the European
Union  Feder funds,  and by  the CIRIT.  L.G.A. also  acknowledges the
Spanish MCYT for a Ram\'on y Cajal Fellowship.
\end{acknowledgements}

\end{document}